\pdfoutput=1
\documentclass[authoryear, 12pt]{elsarticle_Ben}
\usepackage{epstopdf}
\usepackage{graphicx}
\usepackage{amsmath}
\usepackage{hyperref}
\hypersetup{
    colorlinks=true,
    linkcolor=blue,
    filecolor=magenta,
    urlcolor=cyan,
}
\usepackage{lineno}
\usepackage{geometry}
\usepackage{color}
\usepackage{fullpage}
\usepackage{verbatim}
\usepackage{pdfpages}
\usepackage{lscape}
\usepackage{tabulary}
\usepackage{tabularx}
\usepackage{soul}
\sethlcolor{yellow}
\setulcolor{red}
\setstcolor{lightgreen}
\usepackage{natbib}

\usepackage{color}
\definecolor{grey}{rgb}{.6,.6,.6}
\definecolor{dred}{rgb}{.5,0,0}
\definecolor{dgreen}{rgb}{0,.6,0}

\begin{document}
%\linenumbers
\begin{frontmatter}
\title{\bf{A note on the impact of news on US household inflation expectations}\tnoteref{t1}}
\tnotetext[t1]{We thank the editor and the two anonymous referees for their constructive comments. \\ This is a post-peer-review, pre-copyedit version of an article published in Macroeconomic Dynamics. The final authenticated version is available online at: \url{http://dx.doi.org/10.1017/S1365100518000482}}

\author[MQ]{Ben Zhe Wang\corref{cor1}}
%\ead{ben.wang@mq.edu.au}
\author[MQ]{Jeffrey Sheen}
%\ead{jeffrey.sheen@mq.edu.au}
\author[MQ]{Stefan Tr{\"u}ck}
%\ead{stefan.trueck@mq.edu.au}
\author[PD]{Shih-Kang Chao}
%\ead{chaosh@missouri.edu.edu}
\author[HM]{Wolfgang Karl H\"{a}rdle}
%\ead{stat@wiwi.hu-berlin.de}
\address[MQ]{Macquarie University}
%\address[MQ]{Department of Actuarial Studies and Business Analytics, Macquarie University}
\address[HM]{Humboldt-Universit\"{a}t zu Berlin and Sim Kee Boon Institute for Financial Economics, Singapore Management University}
\address[PD]{University of Missouri}
\cortext[cor1]{Corresponding author, Email: ben.wang@mq.edu.au}
%\begin{abstract}
%\doublespacing
%Monthly disaggregated US data from 1978 to 2016 reveals that exposure to news on inflation and monetary policy helps to explain inflation expectations. This remains true when controlling for household personal characteristics, perceptions of government policy effectiveness, future interest rates and unemployment expectations, and sentiment. We find an asymmetric impact of news on inflation and monetary policy after 1983, with news on rising inflation and easier monetary policy having a stronger effect in comparison to news on lowering inflation and tightening monetary policy. Our results indicate the impact on inflation expectations of monetary policy news manifested through consumer sentiment during the lower bound period.
%\textcolor[rgb]{1.00,0.00,0.00}{After 2008, news on monetary policy operating through credit market easing provides a signal for reducing inflation expectations, conditioned by a declining belief in policy effectiveness and forward guidance from the FED.}
%\end{abstract}
%\begin{keyword}
%Inflation expectations \sep news impact \sep monetary policy signalling \sep unconventional monetary policy \\
%JEL Classification: C81, D83, D84, E31\\
%\end{keyword}
\end{frontmatter}

%%%%%%%%%%%%%%%%%%%%%%%%%%%%%%%%%%%%%%%%%%%%%%%%%%%%%%%%%%%%%%%%%%%%%%%%%%%%%%%%%%%%%%%%%%%%%%%%%%%%%%%%%%%%%%%%%%%%%%%%%%%%%%%%%%%%%%%%%%%%%%%%%%%%%%%%%%%%%%%%%%%%%%%%%%%%%%%%%%%%%%%%%%%%%%%%%%%%%%%%%%%%%%%%%%%%%%%%%%%%%%%%%%%%%%%%
%%%%%%%%%%%%%%%%%%%%%%%%%%%%%%%%%%%%%%%%%%%%%%%%%%%%%%%%%%%%%%%%%%%%%%%%%%%%%%%%%%%%%%%%%%%%%%%%%%%%%%%%%%%%%%%%%%%%
\newpage
{\bf Running title:}\\

News Impact on US inflation expectations \\

{\bf Corresponding author:}\\

Name: Ben Zhe Wang\\
Email: ben.wang@mq.edu.au\\
Tel: +61 2 98508500
Address: 4ER 432, Department of Economics, Macquarie University, North Ryde, 2109, NSW, Australia

\newpage

{\bf Abstract}\\

Monthly disaggregated US data from 1978 to 2016 reveals that exposure to news on inflation and monetary policy helps to explain inflation expectations. This remains true when controlling for household personal characteristics, perceptions of government policy effectiveness, future interest rates and unemployment expectations, and sentiment. We find an asymmetric impact of news on inflation and monetary policy after 1983, with news on rising inflation and easier monetary policy having a stronger effect in comparison to news on lowering inflation and tightening monetary policy. Our results indicate the impact on inflation expectations of monetary policy news manifested through consumer sentiment during the lower bound period.\\

{\bf Keywords}\\

Inflation expectations; news impact; monetary policy signalling; unconventional monetary policy \\

{\bf JEL Classification: }\\

C81, D83, D84, E31\\

%\begin{comment}
%\doublespacing
\newpage
\section{Introduction}
Inflation expectations play a major role in modern macroeconomics, with rational expectations ubiquitous as the modelling device for a representative agent.  However, the literature provides both theoretical models and empirical observations that can explain how different economic agents form inflation expectations and why they might disagree on their forecasts. For example, \cite{MankiwReisWolfers2004} document a considerable degree of disagreement in surveys of US inflation expectations. This disagreement is time-varying and exhibits covariation with macroeconomic variables. \cite{MankiwReis2002} construct a formal  model and attribute disagreements to information rigidity. The idea is that the dissemination of new information occurs gradually between people. \\

One way households acquire information is through media reports, which we refer to as `news' in this paper. News can directly impact on household inflation expectations by directly informing the consumer about the possible future path of inflation (e.g. through expert forecasts), or indirectly through impacting on household perceptions of current inflation. \cite{LamlaMaag2012} find that the disagreement in household inflation expectations in Europe depends on the reporting intensity and the `tone' of the news about inflation, while \cite{Drager2015} finds that the media has a small but significant impact on inflation expectations in Sweden.  \cite{Carroll2003} uses an epidemiology model and finds that professional forecasts as a proxy for news have predictive power for household forecasts in the US.\\

All the aforementioned studies use aggregated news measures obtained from a separate source than that for the measure of inflation expectations. One drawback with this approach is that the news measures do not necessary reflect the news heard by the individual household, and thus may not necessarily be attributable to household inflation expectation formation. In this paper, we use the Michigan Survey of Consumers from 1978 to 2016, which allows us to examine the direct impact of news on individual households.\\

There is an emerging literature on investigating the effect of perceived news using the Michigan Survey of Consumers data. For example, utilising the panel structure of the Michigan Survey of Consumer data,\footnote{Each month, about 40 per cent of the households are randomly chosen to be re-interviewed six months after their initial interview. This rotating panel feature is useful for analysing how consumers update their inflation expectations.}, \cite{PfajfarSantoro2013} test the epidemiology model of \cite{Carroll2003} using an aggregate measure of news and household perceived news, and find at best weak support for the epidemiology model. Although hearing inflation news increases the probability of updating inflation expectations, it enlarges the forecast gap between households' inflation expectation and those of professional forecasts, as well as the gap between households' inflation expectation and actual realized inflation. Similarly, \cite{DragerLamla2017} find the hearing of news on inflation increases the chance of households updating their inflation expectations, irrespective of whether it is favourable or unfavourable news. \cite{PfajfarSantoro2009} find households with different socioeconomic background form inflation expectation differently in response to inflation news, and they exhibit different degrees of information stickiness when updating their inflation expectations. In addition, \cite{Ehrmannetal2017} find households tend to forecast inflation higher if they have financial difficulties or are pessimistic about major purchases, income developments or the unemployment rate---however, their bias shrinks by more than the average household in response to inflation news. \cite{Lahiri2016} also find consumer sentiment responds to perceived news and \cite{Zhangetal2016} find stock markets react to news through its impact on sentiment. \\

%We contribute to the literature by exploring the details of the Michigan survey of consumers (MSC) data, directly estimating the news impact on household inflation expectations using this database with a long time series of cross-sectional data. The MSC data presents the answers of respondents concerning their inflation expectations and asking whether they have heard particular news in the past few months, and if they have, to provide the news content. \\

In this paper, we contribute to the literature by considering monetary policy news along with inflation news, and evaluate whether favourable or unfavourable news have asymmetric impacts on household inflation expectations. Unlike \cite{PfajfarSantoro2013} and \cite{DragerLamla2017}, we do not restrict ourselves to using the panel structure of the Michigan data to investigate how inflation expectations are updated according to news, but rather examine how inflation expectations are formed in general.\\

Our results using the Michigan Survey from 1978 to 2016 show that households raise their inflation expectations when they are exposed to news of rising inflation and of contractionary monetary policy. The latter result is an indication that monetary policy acts as a signalling device for the formation of inflation expectations.\footnote{Our paper is also related to a growing theoretical literature that shows monetary policy could have real effects even in the absence of nominal rigidities, if we are willing to not assume rational expectations. The transmission channels may arise from information rigidities \citep{Woodford2001}, rational inattention \citep{Adam2007} and potential signalling effects \citep{Melosi2017}. } Our results are robust after controlling for household demographics, their perception of the effectiveness of government policies, their expectations of future interest and unemployment rates, and their sentiment. We also find an asymmetric impact of news on rising inflation (contractionary monetary policy) compared to news on falling inflation (expansionary monetary policy). Our results indicate that this asymmetric impact started to become significantly stronger in the early 1990s. We find that the absolute impact of news on higher inflation became statistically greater than news on lower inflation after 1991, while after 1999 news on easing monetary policy had a significantly greater impact on inflation expectations than contractionary monetary policy. Finally, during the zero lower bound period after 2008, news about  monetary policy becomes an imperfect signal for inflation expectations formation. This signal manifested through consumer sentiment, which implies central banks should pay attention to consumer sentiment when communicating monetary policies.\\

The subsequent paper is organized as follows. Section \ref{sec:Model} describes the applied model and the data used. Section \ref{sec:News} examines the impact of news on households inflation expectations. Section \ref{sec:Asymm} tests if the content of news has an asymmetric impact on inflation expectations, while Section \ref{sec:ZLB} examine the news effects during the zero lower bound period. Section \ref{sec:Conclusion} concludes.

%%%%%%%%%%%%%%%%%%%%%%%%%%%%%%%%%%%%%%%%%%%%%%%%%%%%%%%%%%%%%%%%%%%%%%%%%%%%%%%%%%%%%%%%%%%%%%%%%%%%%%%%%%%%%%%%%%%%
\section{The model and data} \label{sec:Model}
Since 1978, around 500 adults in households have been surveyed each month on their one-year-ahead inflation expectations by the University of Michigan (Survey of Consumers). The survey asks respondents to provide a numerical answer to the following question:

\begin{quotation}
By about what percent do you expect prices to go (up/down) on the average, during the next 12 months?
\end{quotation}

 The data exhibits a considerable degree of disagreement among these US households in any month. In addition to inflation expectations, the survey also asks respondents whether they have heard news about current economic conditions, and also for their evaluations of current and expected future paths of the economy as well as their personal financial situation. \\

We test if news plays a role in explaining household inflation expectations by estimating equation \eqref{eqn:main} using pooled ordinary least squares (OLS):
  \begin{equation}
    \pi^e_{it} = \alpha + TD_t\theta' + \phi^{\pi} N_{it}^{\pi} +\phi^{r} N_{it}^{r}+  C_{it}\gamma' + \epsilon_{it} \label{eqn:main}
  \end{equation}

where $\pi^e_{it}$ is the one-year ahead inflation expectation of household $i$ at time $t$, $\alpha$ is a constant, and $TD_t$ collects monthly time dummies that are invariant among households at a given month. Since our focus is on investigating the impact of news on individual inflation expectations, we include these time dummies to account for aggregate developments of the economy in each month that might have an impact on household inflation expectations. \\

$N_{it}^{\pi}$ and $N_{it}^{r}$ indicate whether household $i$ has been exposed to any news of inflation and monetary policy, respectively. The survey asks respondents to indicate whether they have heard news of changes in business conditions:

\begin{quotation}
During the last few months, have you heard of any favorable or unfavorable changes in business conditions? What did you hear?
\end{quotation}

The respondents may indicate they have heard news on rising or falling prices, which we use to approximate inflation news, and lower or higher interest rates or easier or tighter credit conditions, which we use to approximate monetary policy news.\\

If no particular news has been heard, the respective variable has a value of 0; $N_{it}^{\pi}$ is set to a value of $1$ if household $i$ has been exposed to news about higher inflation, and $-1$ in the case of news about lower inflation. In the same manner, $N_{it}^{r}$ takes on a value of $1$ if household $i$ has heard news about higher interest rates or tighter credit conditions, mostly associated with tighter monetary policy, and $-1$ for exposure to news about lower interest rates or easier credit conditions--expansionary monetary policy. $\phi^{\pi}$ and $\phi^{r}$ measure the impact of inflation news and monetary policy news on household inflation expectations, which is a key focus of this paper. \\

$C_{it} \in [D_{it}, P_{it}, E_{it}, CS_{it}]$ represents control variables for the characteristics of household $i$. Hereby, $D_{it}$ denotes economic and demographic variables for respondent $i$, including log income, age, gender (1 for a female), and level of education (measured on a scale between 1 to 6, with 6 indicating the highest level of education).\footnote{A value of 1 indicates Grade 0-8 without high school diploma; 2 indicates grade 9-12 without high school diploma; 3 indicates grade 0-12 with high school diploma; 4 indicates grades 13-17 without a college degree; 5 indicates grade 13-16 with a degree and 6 indicates grade 17 with a college degree.} $P_{it}$ denotes household perceptions on the effectiveness of government policies in managing inflation or unemployment, taking a value of 1 if the government is perceived to have done a good job, 0 for a fair job, and $-1$ for a poor job.\footnote{The survey asks respondents to provide their opinion on the following question: \begin{quotation} As to the economic policy of the government -- I mean steps taken to fight inflation or unemployment -- would you say the government is doing a good job, only fair, or a poor job?\end{quotation}} $E_{it}$ collects household expectations on the future course of interest rates and unemployment over the next year, with 1 indicating that household $i$ expects the future respective rate will increase, 0 that it stays the same, and -1 indicates that the household expects the rate to be reduced.\footnote{The survey asks respondents to provide their forecast of interest rate and unemployment: \begin{quotation} No one can say for sure, but what do you think will happen to interest rates for borrowing monetary during the next 12 months--will they go up, stay the same, or go down?\end{quotation} \begin{quotation} How about people out of work during the coming 12 months--do you think that these will be more unemployment than now, about the same or less?\end{quotation}} $CS_{it}$ is the measure from the Michigan Survey for consumer sentiment. It is constructed from five qualitative questions about the household's current and future expected personal financial situation, its current buying attitude regarding large ticket household items, and its expectation of short and medium term business conditions.\footnote{More details about the calculation can be found at: \url{https://data.sca.isr.umich.edu/fetchdoc.php?docid=24770}} \ref{App:Corr} shows the pairwise correlations among the explanatory variables. Apart from log income and education, and the perception of government policies and consumer sentiment, all explanatory variables are not highly correlated.  \\

Our monthly sample starts from January 1978 and ends in February 2016, containing 208,777 individual records in total. The cross-sectional inflation expectations data ranges from 0 to (a cap at) 50 percent. We follow the literature, see e.g. \cite{Curtin1996}, and restrict our sample to those respondents who gave inflation expectations below 30 percent, on the grounds that such outliers are likely to be frivolous. Since our sample covers the high inflation episode in the late 1970s, and there are possible structural breaks in the inflation expectations series, we test for structural breaks in median household inflation expectations following \cite{BaiPerron2003}. The results suggest a structural break in median household inflation expectations in September 1983 so we split our sample into pre- and post-September 1983 periods. \\

    \begin{figure}[h]
      \begin{center}
        \includegraphics[scale=0.75,angle=0]{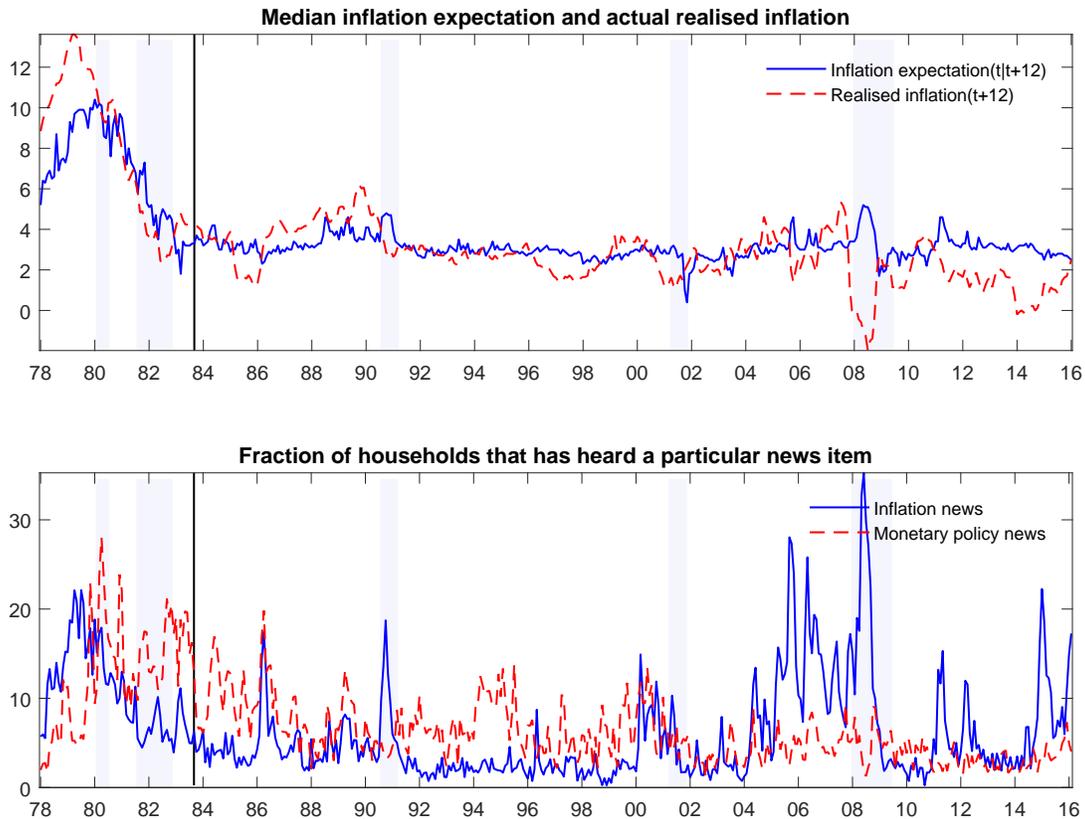}
        \caption{The top panel shows the median 12-month ahead household inflation expectation and the realized 12-month ahead inflation. The bottom panel shows the fraction of households that have heard inflation news or monetary policy news.}
        \label{fig:InfExp}
     \end{center}
    \end{figure}

The top panel of Figure~\ref{fig:InfExp} shows the 12-month ahead median household inflation expectations (blue solid line) and the actual realized 12-month ahead inflation (red dashed line), with shaded areas indicating NBER-dated recessions and the vertical line showing our structural break date. Both actual and expected inflation were high in the late 1970s but gradually decreased during the two recessions in the early 1980s. Both remained relatively low throughout the 1990s and early 2000s.\footnote{The median inflation expectation in the first sub-sample (1978:01-1983:09) was 6 per cent, compared with 3 per cent in the second sub-sample (1983:10-2016:02). This reduction in the median expectation was accompanied by a reduction in the heterogeneity of inflation expectations, with the variance of the cross-sectional distribution decreasing from 34.3 in the first sub-sample to 15.2 in the second sub-sample. This reduction in the heterogeneity of inflation expectations was likely due to the low and stable inflation rate in the second sub-sample, and a stronger emphasis placed by the FED on maintaining low and stable inflation.} It is interesting that households on average also expected higher inflation during and after the financial crisis of 2008. These expectations remained much greater than realized inflation, while deflation was likely more of a concern to policy makers. \\

The bottom panel shows the fraction of households that had heard inflation news (blue solid line) or monetary policy news (red dashed line). As illustrated by the figure, the fraction of households who had heard inflation news is quite volatile. As proposed by \cite{Ehrmannetal2017}, this fraction is often driven by people who have heard that prices are higher. The authors also find the fraction to be highly correlated with retail gasoline price inflation in general, suggesting that frequently purchased items shape households' inflation (news) perceptions. The spikes in the series could be related to economic recessions or actual low or high inflation rates. For example, the high percentage of households who had heard inflation news in the period March-May 1986 can be explained by the fact that in 1986 inflation rates had reached levels below 2\% for the first time in 20 years, a situation that was frequently discussed in the media. The spike in the series in September-November 1990 is most likely a result of the US entering into recession in July 1990, lasting until March 1991. The recession was at least partially related to the restrictive monetary policy enacted by the Federal Reserve throughout 1989 and 1990, when the stated policy was to reduce inflation. The high fraction of households who had heard inflation news in the early and mid 2000s was typically related to news about higher prices. For example, during the first three months of 2004, consumer prices increased at a seasonally adjusted annual rate of over 5\% which was much higher than in previous years. In September 2005, the consumer price index had also risen again by almost 5\% in comparison to 12 months earlier. Notably, a significant part of both increases could be related to rising energy costs, an issue often reported in the news around that time. The fraction of households that had heard news about inflation was also extremely high in the second half of 2008. This is most likely related to the subprime mortgage crisis and the fact that both households and the media paid far more attention to news about macroeconomic and financial conditions during that period.\\

Generally, most of the spikes are due to periods where people had heard news on higher prices, except for three episodes: between March and May 1986, when inflation rates below 2\%; between December 2014 and February 2015, when the US economy deflated for the first time since 2009; and from January 2016 to the end of the sample, when a general discussion on the risks of a prolonged deflation period became more prevalent in the media. It is interesting that news of lower inflation were rare, and became prevalent only near the end of the sample, even though a fear of deflation had become a widespread concern among policy makers and commentators after 2008.\\

Interestingly, the fraction of households that heard monetary policy news has remained low since the 2000s, and this is true even during the recent global financial crisis when the Federal Reserve had used extensively unconventional monetary policies.\footnote{The fraction of household that had heard inflation (monetary policy) news was 11.33(12.47) percent on average for the first sub-sample, decreasing to 5.61(5.96) per cent in the second sub-sample.}\\

%\textcolor[rgb]{1.00,0.00,0.00}{[[[STEFAN: to add some discussion about the spikes in panel 2 in 1986, 91, 2000, 2006,7,8, 11, 13 15]]]}\\
%\footnote{We conducted multiple breakpoint tests using \cite{BaiPerron2003} for the sample period. The LWZ criterion suggested that there was only one break in September 1983. On the other hand, there were two breaks in September 1983 and March 1991 according to the Bayesian Information Criterion (BIC). However the biggest gain by far in the likelihood was for the first break, while the other break was marginal. Therefore we selected one structural break in September 1983.} \\

%%%%%%%%%%%%%%%%%%%%%%%%%%%%%%%%%%%%%%%%%%%%%%%%%%%%%%%%%%%%%%%%%%%%%%%%%%%%%%%%%%%%%%%%%%%%%%%%%%%%%%%%%%%%%%%%%%%%
\section{News impact on household inflation expectations} \label{sec:News}
We test if exposure to direct news of inflation and monetary policy affects household inflation expectations. Model 1 considers only the impact of news on inflation and Model 2 considers both news on inflation and monetary policy without controlling for characteristics of household $i$, thus serving as a benchmark. The subsequent five models extend the benchmark specification: Model 3 controls for the additional impact of demographic characteristics $D_{it}$; Model 4 controls for the additional impact of perceptions of the effectiveness of government policies $P_{it}$; Model 5 controls for the additional impacts of expectations about interest rates and unemployment; Model 6 controls for the additional impacts of consumer sentiment on inflation expectations $ICS_{it}$; and Model 7 considers all explanatory variables jointly. We report results in Table \ref{tab:OLS}, with the top panel of the table showing the results for January 1978 to September 1983, the middle panel focuses on October 1983 to February 2016. Since we have 457 monthly time dummies and their interpretations do not necessarily relate to news effects, we omit results for the time dummies in the table.\footnote{The monthly dummies capture the effect of common factors on household inflation expectation. One of these factors may be the objective intensity of news reporting on inflation and monetary policy.}
\linespread{1}
      \begin{table}[h!]
     {\footnotesize
      \begin{center}
      \begin{tabular}{@{}l@{ }l l l l l l l@{}}
        \hline
        \hline \vspace{-0.2cm} \\
        & \multicolumn{7}{c}{Sub-sample 1: 1978:01 -- 1983:09}  \\
        \cline{2-8}  \vspace{-0.3cm}\\
                   & Model 1 & Model 2 & Model 3 & Model 4 & Model 5 & Model 6 & Model 7\\
                    \cline{2-8}  \vspace{-0.3cm}\\
                               Constant   &   4.45*** &   4.48*** &   4.43***  &   4.56***   &   4.47***  &   6.08***  & 4.04*** \\
          News: inflation($\phi^{\pi}$)   &   1.03*** &   0.98*** &   0.94***  &   0.83***   &   0.83***  &   0.87***  & 0.67*** \\
        News: monetary policy($\phi^r$)   &       -- &   0.49*** &   0.47***  &   0.37***   &   0.35***  &   0.40***  & 0.22*** \\
                             Log income   &       -- &       -- &   0.07     &       --   &       --  &       --  & 0.21*** \\
                                    Age   &       -- &       -- &  -0.03***  &       --   &       --  &       --  &-0.03*** \\
                                 Female   &       -- &       -- &   0.26***  &       --   &       --  &       --  & 0.06    \\
                              Education   &       -- &       -- &   0.15***  &       --   &       --  &       --  & 0.19*** \\
          Perception: government policy   &       -- &       -- &       --  &  -1.08***   &       --  &       --  &-0.73*** \\
             Expectation: interest rate   &       -- &       -- &       --  &       --   &   0.82***  &       --  & 0.72*** \\
         Expectation: unemployment rate   &       -- &       -- &       --  &       --   &   0.82***  &       --  & 0.48*** \\
                     Consumer sentiment   &       -- &       -- &       --  &       --   &       --  &  -0.02***  &-0.01*** \\

        \hline \vspace{-0.4cm} \\
                         Adjusted-$R^2$   &    0.623 &    0.624  &    0.628   &    0.629&    0.633&    0.628&    0.642 \\
  \hline \vspace{-0.2cm}\\
  &\multicolumn{7}{c}{Sub-sample 2: 1983:10 -- 2016:02} \\
  \cline{2-8}  \vspace{-0.3cm}\\
                               Constant   &   3.21*** &   3.19*** &   9.04***  &   3.08***   &   2.92*** &   4.92***  & 8.87*** \\
          News: inflation($\phi^{\pi}$)   &   0.64*** &   0.62*** &   0.61***  &   0.52***   &   0.50*** &   0.43***  & 0.39*** \\
        News: monetary policy($\phi^r$)   &       -- &   0.35*** &   0.32***  &   0.26***   &   0.18*** &   0.17***  & 0.08**  \\
                             Log income   &       -- &       -- &  -0.45***  &       --   &       -- &       --  &-0.35*** \\
                                    Age   &       -- &       -- &  -0.01***  &       --   &       -- &       --  &-0.01*** \\
                                 Female   &       -- &       -- &   0.66***  &       --   &       -- &       --  & 0.54*** \\
                              Education   &       -- &       -- &  -0.14***  &       --   &       -- &       --  &-0.11*** \\
          Perception: government policy   &       -- &       -- &       --  &  -0.75***   &       -- &       --  &-0.33*** \\
             Expectation: interest rate   &       -- &       -- &       --  &       --   &   0.47*** &       --  & 0.40*** \\
         Expectation: unemployment rate   &       -- &       -- &       --  &       --   &   0.73*** &       --  & 0.35*** \\
                     Consumer sentiment   &       -- &       -- &       --  &       --   &       -- &  -0.02***  &-0.01*** \\
           \hline \vspace{-0.4cm} \\
                        Adjusted-$R^2$   &    0.520  &    0.521  &    0.532   &    0.529&    0.532&    0.535&    0.549 \\
        \hline \vspace{-0.3cm} \\
         \multicolumn{8}{l}{Notes: 1. The sample size is 43,599 between January 1978 and September 1983 and 165,178 between October} \\
         \multicolumn{8}{l}{\qquad \qquad \;  1983 and February 2016.} \\
         \multicolumn{8}{l}{\qquad \; \; 2. *, **, *** represent significance at 10\%, 5\% and 1\% levels of significance.} \\
         \hline
         \hline
      \end{tabular}
      \caption{Regression results for inflation expectations}
      \label{tab:OLS}
      \end{center}}
    \end{table}

\linespread{1.6}
\subsection{The first sub-sample---January 1978 to September 1983}

Results for the first sub-sample show news of inflation and monetary policy having a strong impact on household expectations. In this relatively high inflation period, Model 1 shows that hearing news of higher inflation led an average household to increase their inflation expectations by 1.03 per cent. When jointly considered with news of monetary policy, the impact of news of inflation reduces to 0.98. \\

News of monetary policy changes can affect inflation expectations in two opposite ways. One is if households understand the \emph{transmission mechanism} of monetary policy to future inflation, in which case news of tighter monetary policy implies lower expected future inflation. The other is if households do not understand the transmission mechanism but understand that the central bank targets inflation, in which case news of contractionary monetary policy is a \emph{signal} that inflation is higher than previously expected. Model 2 shows that hearing news of contractionary monetary policy induced households to expect 0.49 per cent higher inflation, indicating households understood that higher interest rates are a result of the central bank's concern about higher future inflation\footnote{A potential endogeneity problem arises for the two types of news used in this paper. For example, in response to tighter monetary policy there may be a perception in this news that inflation prospects will be mitigated. To address this issue, we re-ran the estimation excluding those households that indicated they have heard news on lower inflation and tightening monetary policy and those households that indicated they have heard news on higher inflation and easing monetary policy. There are only marginal changes to the estimated coefficients and all results remain qualitatively the same. The results are available from the authors upon request.}. Therefore the average household expectations \emph{appear} to be informed as a signal by the central bank's response function, for example a Taylor rule (as suggested by \cite{CarvalhoNechio2014}), rather than households concerning themselves with the expected future contractionary effect of the interest rate change on inflation.\footnote{We cannot rule out the possibility that households form  higher inflation expectations when hearing of contractionary monetary policy because they may think the monetary policy is too accommodative, see, for example, \cite{ClaridaGaliGertler2000} and \cite{Gertleretal1999}. A consensus about accommodative monetary policy contributing to high inflation was achieved much later (in the 1980s) and in the first sample period it was surely not well understood  when households formed their inflation expectation. Therefore we do not expect the \emph{average} household would form inflation expectation in this sophisticated way.}   The results from Model 1 and 2 are consistent with earlier findings using \emph{aggregate} data from the media having a role in driving household inflation expectations \citep{LamlaMaag2012,Drager2015}. Our results are also consistent with the information rigidity hypothesis, which suggests that households' private information sets (through news) play a role in explaining disagreements in inflation expectations \citep{MankiwReis2002,MadeiraZafar2015}.\\

Model 3 confirms results in the earlier literature that households with different demographic backgrounds disagree on inflation expectations, see, for example, \cite{BryanVenkatu2001a, BryanVenkatu2001b} for the US, \cite{Blanchfloweretal2009} for the UK, \cite{Easawetal2013} for Italy, and \cite{Jonung1981} for Sweden. Households with different demographics may purchase different consumption bundles. In particular, we find that those who are younger, female, and better educated tended to forecast higher inflation levels. Households may form inflation expectations according to their lifetime inflation experience. Therefore younger people expected much higher inflation following the years of high inflation in the 1970s. Indeed \cite{MalmendierNagel2016} find households with members older than 70 years expected lower inflation compared to households with members younger than 40 years in the 1970s. It is well documented that women on average forecast higher inflation than men. One possibility is that women have higher perceived inflation (for example \cite{Jonung1981}) due to being likely responsible for grocery shopping and thus exposed more often to prices than men. The impact of education on inflation forecasts is interesting. As seen in Figure~\ref{fig:InfExp}, median inflation expectations underestimate actual inflation in the higher inflation period of the first sub-sample (and overestimate inflation in the lower and more stable inflation period of the second sub-sample). The fact that better educated households forecast higher inflation in the first sub-sample (and lower inflation in the second sub-sample) indicates they are better forecasters than the median household. Better educated households thus appear to have better understood the severity of the implications of oil price shocks on inflation in the 1970s, realizing that monetary policy would need to be strategically accommodative to minimize the effects of the rise in the relative oil price. This accommodative monetary policy did not increase the nominal interest rate more than inflation, thus reducing the real interest rate and making monetary policy expansionary, as argued by \cite{ClaridaGaliGertler2000}. A similar logic may apply to households with higher income, if we assume that these households are more likely to participate in financial markets, and thus tend to be better informed.\\

Perceiving government policies to be effective in managing the business cycle significantly reduced inflation expectations by 1.08 per cent (Model 4). Households that heard tightening monetary policy news and who believe government policies are effective will form 0.71 (0.37-1.08) percent lower inflation expectation compared to the average household, whereas households hearing the same news but who do not believe policies are effective would expect 1.45 (0.37+1.08) per cent higher inflation than the average household. This appears to indicate that households who believe government policies are effective tend to forecast inflation consistent with the transmission mechanism of monetary policy rather than being informed as a signal by monetary policy. This result also implies that if the government wants to lower inflation, it can reduce inflation expectations by influencing household perceptions about the effectiveness of its policies. \\

Expecting a rise in interest rates over the next year was associated with higher inflation expectations (Model 5). This confirms the findings in Model 2 and suggests that households understood monetary policy responds to inflation now and in the near future, so that higher expected inflation is associated with  higher current and expected future interest rates. Also higher expected future unemployment was associated with higher inflation expectations. These results suggest that the average household seemed not to be concerned with the implied negative correlation between expected inflation and expected unemployment of an expectations-augmented Phillips curve, instead associating higher expected inflation with a higher expected future unemployment rate.\\

The negative and significant consumer sentiment parameter estimate (Model 6) shows that more optimistic households expected slightly lower inflation than the average. This result indicates that households' perception and sentiment---reflecting their interpretation of their private information set---help to explain why they disagreed in their inflation expectations. This parameter estimate is robust to the sample used.  \\

%Though news impacts remain economically and statistically significant, adding perception, expectations and sentiment lowers these impacts. This suggests that news affected household inflation expectations partly through changes in household perceptions of the effectiveness of government policies, expectations on the future course of interest rate and unemployment rate and their sentiment levels. \\

Model 7 includes all regressors and shows that the impact of news about inflation and monetary policy, perceptions on the effectiveness of government policies, expectations of future interest rates and the unemployment rate as well as consumer sentiment were all important factors for explaining the heterogeneity of inflation expectations. Adding perceptions on government policies, expectations and sentiment induced a magnification of the impacts of income, while gender had a lessened impact. Therefore in the first sub-sample, those who were richer tended to expect higher inflation, owing to their perceptions, expectations and sentiment being less positive than those of poorer households.\\

 \subsection{The second sub-sample---October 1983 to February 2016}
Many of the aforementioned results remain true in the second sub-sample. But there are some notable differences. First, the impacts on inflation expectations of news about both inflation and monetary policy were smaller, although they remain significant. This lower impact of news in the second sub-sample reflects the fact that inflation had fallen and stabilised during this period, making news of inflation and monetary policy less salient for households, and thus reducing their impact on inflation expectations. Second, the signs of the impact of household income changed to be negative, and was much larger in absolute size. Third, the sign on education also reversed so that now better than average educated households expected lower inflation. These two sign reversals mean that households with higher income and better education forecasted lower inflation than the average household in this sub-period. Finally, gender played a much larger role, with the difference between male and female expectations becoming significantly larger.

%\textcolor[rgb]{1.00,0.00,0.00}{Finally, while consumer sentiment continues to have the same negative effect, in section \ref{sec:ZLB} where we focus on the zero lower bound period from 2008, it appears to play a particularly strong role, which will be discussed there.}\\

In summary over the whole sample, our results indicate that exposure to news of higher inflation and contractionary monetary policy significantly increased household inflation expectations. This result is robust across sample periods, and holds even after controlling for household demographic characteristics, their perceptions on the effectiveness of government policies, their expectations about interest rates and unemployment, and their sentiment. Among macroeconomic theories, information rigidity models have been widely used to explain cross-sectional disagreements of inflation expectations, see, for example \cite{MankiwReis2002} and \cite{MankiwReisWolfers2004}. These models typically assume information is costly to acquire, so people have a different information set when forecasting future paths of the economy. Our results indicate that households who had a larger news exposure expected different inflation rates (\emph{ceteris paribus}), thus supporting the information rigidity theory. The fact that the estimated news effect ($\phi^{\pi}$ and $\phi^{r}$) between Model 2 and 3 are very similar suggests that household demographics and news almost independently explain inflation expectations. This means that the demographic impacts on inflation expectations were not due to the different demographic groups' exposure to news. \\

Controlling for the perception of the effectiveness of government policies, for expectations on future interest and unemployment rates and for consumer sentiment reduces the impact of news on household inflation expectations. These findings suggest that news of inflation and monetary policy impacted on inflation expectations partially through these household perceptions about policy effectiveness, their expectations of future interest rates and unemployment, and their sentiments about current economic conditions.

%%%%%%%%%%%%%%%%%%%%%%%%%%%%%%%%%%%%%%%%%%%%%%%%%%%%%%%%%%%%%%%%%%%%%%%%%%%%%%%%%%%%%%%%%%%%%%%%%%%%%%%%%%%%%%%%%%%%
\section{The asymmetric impact of news} \label{sec:Asymm}
News on the movements of underlying economic variables may have an asymmetric impact on inflation expectations. This may arise if one particular direction of movement of the variable has a more salient effect on expectations than the other at the time of making the expectation decision. For example, due to diminishing marginal utility, higher inflation can erode household wealth and reduce utility more than it would increase it, if inflation fell by the same amount--households may thus pay more attention to news of higher inflation than of lower inflation. Households may also have experienced the high inflation episodes in the 1970s and understand high inflation may indicate unsuccessful policies and have long lasting effects on future paths of inflation (\cite{MadeiraZafar2015}) compared to lower inflation. Thus, it may be reasonable to assume a bigger impact of high inflation on future inflation expectations. Using aggregated data, \cite{LamlaMaag2012} and \cite{Drager2015} indeed find the content of media reports have an asymmetric impact on inflation expectations.  \\

Utilising the cross-sectional nature of the MSC inflation expectations and news data, we investigate whether news content has an asymmetric impact on household inflation expectations at the disaggregated level. For each of the news variables $N_{it}^{\pi}$ and $N_{it}^{r}$ considered, we construct two dummy variables according to the content of the news.  An upward arrow $\uparrow$ denotes news that corresponds to an increasing value of the underlying variable, while a downward arrow $\downarrow$ relates to news decreasing the value of the underlying variable. For example, news of rising inflation would result in a value of 1 for $N_{it}^{\pi}\uparrow$ and a value 0 for $N_{it}^{\pi}\downarrow$; $N_{it}^{r}\downarrow=1$ indicates news about easing monetary policy and $N_{it}^{r}\uparrow = 1$ indicates news about contractionary monetary policy. We thus replace $N_{it}^{\pi}$ and $N_{it}^{r}$ in Equation \eqref{eqn:main} with $N_{it}^{\pi}\downarrow, N_{it}^{r}\downarrow$ and $N_{it}^{\pi}\uparrow,  N_{it}^{r}\uparrow$:
  \begin{equation}
    \pi^e_{it} =  \alpha  + \phi^{\pi}_{\downarrow} N_{it}^{\pi}\downarrow+ \phi^{\pi}_{\uparrow} N_{it}^{\pi}\uparrow +\phi^{r}_{\downarrow} N_{it}^{r}\downarrow +\phi^{r}_{\uparrow} N_{it}^{r}\uparrow +  C_{it}\gamma' + TD_t\theta' +  \epsilon_{it} \label{eqn:main_Assym}
  \end{equation}

For $j \in \{\pi, r \}$, we expect $\phi^j_{\downarrow}$ to have the opposite impact on inflation expectations to  $\phi^j_{\uparrow}$. We are interested in testing whether or not increases and decreases have the same absolute impact on inflation expectations. To test this, we calculate the $z$-score between the two estimated parameters and test whether the null hypothesis $ \phi^j_{\downarrow} = - \phi^j_{\uparrow}$ is rejected:
%\begin{equation}
%  z^j = \frac{\phi^j_{\downarrow} - (- \phi^j_{\uparrow})}{\sqrt{\text{Var}(\phi^j_{\downarrow} - (- \phi^j_{\uparrow}))}} = \frac{\phi^j_{\downarrow} + \phi^j_{\uparrow}}{\sqrt{\text{Var}(\phi^j_{\downarrow}) + \text{Var}(\phi^j_{\uparrow}) + 2\text{Cov}(\phi^j_{\downarrow}\phi^j_{\uparrow})}} \label{eq:Assymtest}
%\end{equation}
\begin{equation}
  z^j = \frac{\phi^j_{\downarrow} - (- \phi^j_{\uparrow})}{\sqrt{\text{Var}(\phi^j_{\downarrow} - (- \phi^j_{\uparrow}))}} \label{eq:Assymtest}
\end{equation}
\linespread{1}
  \begin{table}[h!]
     {\footnotesize
      \begin{center}
      \begin{tabular}{l c c}
        \hline
        \hline \vspace{-0.2cm} \\
         & Sub-sample 1 & Sub-sample 2\\
         & 1978:01 -- 1983:09 & 1983:10 -- 2016:02 \\
         \hline \vspace{-0.3cm}\\
                                                         Constant   &   4.07***  &   8.84***  \\
                 News: lower inflation($\phi^{\pi}_{\downarrow}$)   &  -0.49**   &  -0.22***  \\
                  News: higher inflation($\phi^{\pi}_{\uparrow}$)   &   0.70***  &   0.46***  \\
              News: easing monetary policy($\phi^r_{\downarrow}$)   &  -0.08     &  -0.15***  \\
            News: tightening monetary policy($\phi^r_{\uparrow}$)   &   0.31***  &  -0.00     \\
                                    Perception: government policy   &  -0.73***  &  -0.33***  \\
                                       Expectation: interest rate   &   0.72***  &   0.40***  \\
                                   Expectation: unemployment rate   &   0.48***  &   0.35***  \\
                                               Consumer sentiment   &  -0.01***  &  -0.01***  \\
                                                       Log income   &   0.20***  &  -0.35***  \\
                                                              Age   &  -0.03***  &  -0.01***  \\
                                                           Female   &   0.07     &   0.54***  \\
                                                        Education   &   0.19***  &  -0.11***  \\
\hline \vspace{-0.4cm} \\
                                                   Adjusted-$R^2$   &    0.642  &    0.549   \\
   Hypothesis: $\phi^{\pi}_{\downarrow} = -\phi^{\pi}_{\uparrow}$ &     0.95 &  2.72*** \\
           Hypothesis: $\phi^r_{\downarrow} = -\phi^r_{\uparrow}$ &     1.37 &   -1.96* \\
                                      \hline \vspace{-0.3cm} \\
         \multicolumn{3}{l}{Notes: 1. Sub-sample 1 is from January 1978 to September 1983, with sample size  43,599.} \\
         \multicolumn{3}{l}{\qquad \quad \, Sub-sample 2 is between October 1983 and February 2016, with sample size 165,178.} \\
         \multicolumn{3}{l}{\qquad \; \; 2. *, **, *** represent significance at 10\%, 5\% and 1\% levels of significance.} \\
         \multicolumn{3}{l}{\qquad \; \; 3. The estimation results for the monthly time dummy are omitted from the table.} \\
        \hline
         \hline
      \end{tabular}
      \caption{Regression results for asymmetric news impacts}
      \label{tab:OLSASYM}
      \end{center}}
    \end{table}
    
\linespread{1.6}
Table \ref{tab:OLSASYM} shows the estimation results of equation \eqref{eqn:main_Assym} with the columns giving the estimation results for the two sub-samples. The results are broadly consistent with those presented in Table~\ref{tab:OLS}. Consistent with our expectations, the two directions of news content had opposite effects on household inflation expectations. This result is robust across both inflation and monetary policy news and across sample periods. \\

Both news of rising and declining inflation had significant impacts on household inflation expectations across both sub-samples. Hearing news of higher inflation in the first sub-sample increased inflation expectations by 0.70 per cent on average, and hearing news of lower inflation reduced inflation expectations by 0.49 per cent on average in this high inflation period. Hearing news on higher inflation in the second sub-sample increased household inflation expectations by 0.46 per cent on average, but being exposed to news on lower inflation only reduced inflation expectations by 0.22 per cent. This result indicates that households respond to news on higher inflation more than to news on lower inflation in general, though the effect of inflation news was much weaker in the second sub-sample. This is especially true for news on lower inflation, where the impact is more than halved in the second sub-sample. The second row of the lower panel of Table \ref{tab:OLSASYM} shows the the significance of the z-score test of equation \eqref{eq:Assymtest} in terms of inflation news ($ \phi^{\pi}_{\downarrow} = - \phi^{\pi}_{\uparrow}$). Consistent with previous results, the test indicates that the symmetric effect of news on inflation cannot be rejected for the first sub-sample period, while the effect became significantly asymmetric in the second sub-sample, where news of higher inflation had a bigger absolute impact than news of lower inflation.\\

News on easing monetary policy did not significantly alter household inflation expectations in the high inflation period (sub-sample 1), but significantly reduced inflation expectations in the second sub-sample. On the other hand, news of tightening monetary policy significantly increased household inflation forecasts in the first sub-sample but was irrelevant in the second sub-sample.\footnote{Our credibility interpretation remains valid after distinguishing easing and tightening of monetary policy news in the second sub-period. Though the average household reduces inflation expectations when hearing news on easing monetary policy, those who perceive effective government policies understand the implication of monetary policy and forecast higher inflation. The detailed results on this are available on request.} The third row of the lower panel of Table \ref{tab:OLSASYM} shows the significance of the z-score test of $ \phi^{r}_{\downarrow} = - \phi^{r}_{\uparrow}$: we find that the symmetric effect of news on monetary policy could not be rejected for the first sub-sample, even though only tightening monetary policy was significant. However news on monetary policy became asymmetric in the second sub-sample, when easing monetary policy had a much bigger absolute impact on inflation expectations than tightening monetary policy.\\

    \begin{figure}[h]
      \begin{center}
        \includegraphics[scale=0.75,angle=0]{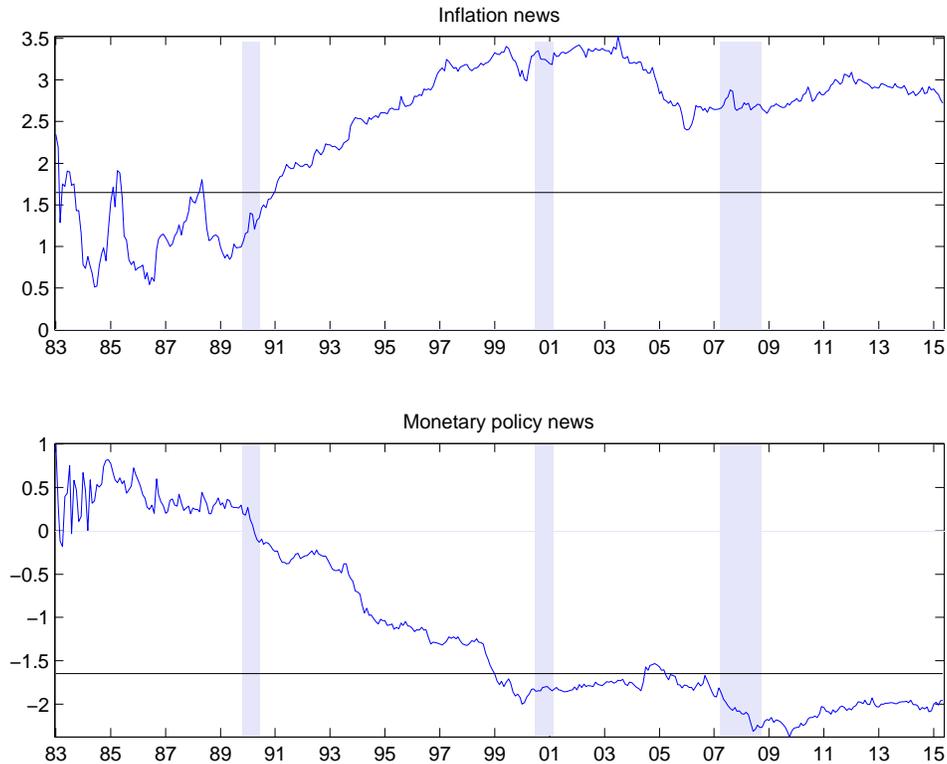}
        \caption{Time-varying asymmetries---z-score tests.}
        \label{fig:Assym_TV}
     \end{center}
    \end{figure}
Since the asymmetries became significant in the second sub-sample, we are interested to know how they evolved over time. We do this by conducting an expanding window estimation of $z^j$ (equation \eqref{eq:Assymtest}) starting in October 1983. Figure \ref{fig:Assym_TV} shows the evolution of $z^j$ for both inflation and monetary policy news, with the horizontal black line indicating significance at the 10 per cent level. It is interesting that both news of increase and decrease on inflation (monetary policy) had similar absolute impacts on household inflation expectations for most of the 1980s. However, both news on inflation and monetary policy started to become increasingly asymmetric in the early 1990s, with the absolute impact of news on higher inflation becoming statistically greater than news on lower inflation after 1991 (top panel of Figure \ref{fig:Assym_TV}), and news on easing monetary policy having a greater impact than contractionary monetary policy after 1999 (bottom panel).\\

One explanation for this interesting evolution of asymmetric news may be rational inattention to information \citep{Sims2003}. Since information is costly to process, households may only pay attention to news information that they regarded as relatively important. A general consensus developed in the 1980s and 1990s was that high inflation was bad and needed to be avoided. Presumably then, high inflation news came to represent unfavourable information for households. As a consequence, low and stable inflation became a norm in the late 1980s and households inflation expectations became firmly anchored around 3 per cent. Even though inflation became a lesser concern, household paid disproportionate attention to news on higher inflation that was regarded as unfavourable. Households may also consider that higher inflation (above the norm) tends to be more persistent compared to lower inflation, thus regarding higher inflation as unfavourable. After 2008, however, there may well have been a growing relative unease about the risks of deflation, but we see no evidence of that in Figure \ref{fig:Assym_TV}, since the z-score tests remained flat. With regards to monetary policy, we find that there is consistently significant evidence since 2007 of a greater impact of news on easing monetary policy in comparison to news on contractionary monetary policy. One interpretation of this result could be that news on upcoming cuts in the federal funds rate (that started to occur in late 2007) as well as on quantitative easing by the Federal Reserve (that started in November 2008) had a noticeably strong impact on inflation expectations.      \\

In summary, we find evidence that rising inflation news and easing monetary policy impacts on household inflation expectations significantly more than does lower inflation and tightening monetary policy. This is true in particular for the relatively lower inflation period (sub-sample 2: 1983:10-2016:02). Extending window estimation shows that the impact of news on higher inflation (easing monetary policy) increasingly became bigger compared to lower inflation (contractionary monetary policy) during the 1990s. These asymmetries on both news persisted through the remaining sample.

%%%%%%%%%%%%%%%%%%%%%%%%%%%%%%%%%%%%%%%%%%%%%%%%%%%%%%%%%%%%%%%%%%%%%%%%%%%%%%%%%%%%%%%%%%%%%%%%%%%%%%%%%%%%%%%%%%%%
\section{The impact of news under the zero lower bound} \label{sec:ZLB}
\linespread{1}
      \begin{table}[h!]
     {\footnotesize
      \begin{center}
      \begin{tabular}{@{}l@{ }l l l l l l l@{}}
        \hline
        \hline \vspace{-0.2cm} \\
        & \multicolumn{7}{c}{Sample period: 2008:06 -- 2016:02}  \\
        \cline{2-8}  \vspace{-0.3cm}\\
                   & Model 1 & Model 2 & Model 3 & Model 4 & Model 5 & Model 6 & Model 7\\
  \cline{2-8}  \vspace{-0.3cm}\\
                               Constant   &   4.04*** &   4.04*** &  10.30***  &   3.79***   &   3.80*** &   5.84***  &10.53*** \\
          News: inflation($\phi^{\pi}$)   &   1.22*** &   1.21*** &   1.08***  &   1.01***   &   0.95*** &   0.80***  & 0.67*** \\
        News: monetary policy($\phi^r$)   &       -- &   0.52*** &   0.52***  &   0.35***   &   0.20*   &   0.14     & 0.07    \\
                             Log income   &       -- &       -- &  -0.49***  &       --   &       -- &       --  &-0.44*** \\
                                    Age   &       -- &       -- &  -0.00     &       --   &       -- &       --  &-0.01*** \\
                                 Female   &       -- &       -- &   0.58***  &       --   &       -- &       --  & 0.48*** \\
                              Education   &       -- &       -- &  -0.26***  &       --   &       -- &       --  &-0.17*** \\
          Perception: government policy   &       -- &       -- &       --  &  -0.90***   &       -- &       --  &-0.37*** \\
             Expectation: interest rate   &       -- &       -- &       --  &       --   &   0.35*** &       --  & 0.35*** \\
         Expectation: unemployment rate   &       -- &       -- &       --  &       --   &   1.14*** &       --  & 0.58*** \\
                     Consumer sentiment   &       -- &       -- &       --  &       --   &       -- &  -0.02***  &-0.01*** \\
           \hline \vspace{-0.4cm} \\
                         Adjusted-$R^2$   &    0.527  &    0.527  &    0.543   &    0.540&    0.549&    0.554&    0.572 \\
        \hline \vspace{-0.3cm} \\
         \multicolumn{8}{l}{Notes: 1. The sample size is 38128 between June 2008 and February 2016.} \\
         \multicolumn{8}{l}{\qquad \; \; 2. *, **, *** represent significance at 10\%, 5\% and 1\% levels of significance.} \\
         \hline
         \hline
      \end{tabular}
      \caption{Regression results for inflation expectations}
      \label{tab:OLS1}
      \end{center}}
    \end{table}

\linespread{1.6}
Does the impact of inflation and monetary policy news change under the zero lower bound from 2008? Table~\ref{tab:OLS1} shows the estimates of Model 1-7 for the period between June 2008 and February 2016. Compared with the second sub-sample in Table~\ref{tab:OLS}, inflation news has a much bigger impact on household inflation expectations for all models. This may reflect the fact that households in this period realize that the FED had lost the effectiveness of its conventional instrument (the federal funds rate) in managing inflation (deflation). Therefore, in this period households may have reacted more sensitively to any news on inflation. Looking again at Figure \ref{fig:InfExp}, note that median inflation expectations were almost always greater than realized inflation from 2008.\\

During this period, the FED could not cut the \emph{current} federal funds rate any further, though it was able to and did use extensively forward guidance, aiming to influence expectations of \emph{future} interest rates and inflation to try to stimulate the economy. Forward guidance can be either Odyssean---when the Feb publicly commits monetary policy to a future action, or Delphic---when the policy states the likely future policy actions based on the policymaker's potential private information about macroeconomic fundamentals, see, for example, \cite{Campbelletal2012}. In addition, the FED undertook 3 rounds of large scale asset purchases from 2008 to 2014, otherwise known as ``quantitative easing", leading to a significant expansion of its balance sheet with bank debt, Treasury securities and mortgage-backed securities.\\

%The effectiveness of such unconventional policies relies crucially on how those economic agents hearing the news of such policies respond. \textcolor[rgb]{1.00,0.00,0.00}{Compared with the second sub-sample of Table~\ref{tab:OLS}, here the negative impact of expansionary monetary policy news on expected inflation \emph{appears} strengthened (Model 2-3), even when the households' demographic backgrounds are jointly considered, thus seeming to strengthen the signalling role of the FED's policy. } \\

Similar to the earlier results, the effectiveness of such unconventional policies relies crucially on how those economic agents respond on hearing the news of these policies. However, compared with the second sub-sample of Table~\ref{tab:OLS}, there are two noticeable differences. First,the impact of monetary policy news on expected inflation \emph{appears} strengthened (Model 2-3), even when the households' demographic backgrounds are jointly considered, thus seeming to strengthen the signalling role of the FED's policy. Second, jointly considering consumer sentiment (Model 6) makes the impact of monetary policy small and statistically insignificant. The comparison of this to the results in Table~\ref{tab:OLS} is indicative of the different implications of unconventional and conventional monetary policy news.  The (unconventional) monetary policy news estimate of Model 2 in Table~\ref{tab:OLS1} may be seen as a proxy for consumer sentiment in relation to inflation expectations formation.Regressing consumer sentiment on monetary policy news yields a significant coefficient (at the 1\% level) of -7.56. Therefore, hearing news on monetary policy contraction would be associated with a 7.56 reduction in consumer sentiment during the zero lower bound period. Consumer sentiment fell significantly in 2008 to 2009, but improved consistently thereafter. The significant negative estimate of consumer sentiment in Model 6 suggests that those households hearing news on monetary policy easing and thus credit easing, recognised this as a signal to lower their inflation expectations and to expect easier conditions that improved consumer sentiment. Households not hearing this news had no signal, and were responsible for maintaining inflation expectations above realized inflation.  \\

In summary, these results suggest in a consistent way that monetary policy news provided a signal about future inflation. This signalling effect manifests through consumer confidence during the zero lower bound period. This implies that central banks should pay particular attention to the impact of their policy communications on consumer sentiment to maximise the impact of asset purchases and forward guidance on inflation expectations. \footnote{We thank the referee for suggesting this.} \\

%%%%%%%%%%%%%%%%%%%%%%%%%%%%%%%%%%%%%%%%%%%%%%%%%%%%%%%%%%%%%%%%%%%%%%%%%%%%%%%%%%%%%%%%%%%%%%%%%%%%%%%%%%%%%%%%%%%%
\section{Conclusions} \label{sec:Conclusion}

We have examined the impact of news on household inflation expectations. Using monthly US consumer inflation expectations data between January 1978 and February 2016, we find that, in general, exposure to news on inflation and monetary policy significantly helps to explain household inflation expectations. This remains true even after controlling for households demographic characteristics, their perception of the effectiveness of government policies in managing business cycles, their expectations of future interest and unemployment rates, and their sentiment. This result tells us that the average effect of news is unaffected by the controls. To understand better other distributional aspects of the response, we would need to consider empirical non-linearities, which we leave for future research.    \\

We find evidence that news on inflation and monetary policy had an asymmetric impact on household inflation expectations.  In particular, households responded to news of higher inflation and easing monetary policy significantly more than news of lower inflation and tightening monetary policy. This was especially true in the relatively low inflation period after 1983, and probably was a result of the broad persuasion by public figures about the dangers of high inflation. The more unfavourable perception of risks of higher inflation remained valid also after 2008, even though, also, the impact of a deflation threat has likely increased since then. If this deflation threat were to become more relevant and households did not yet realize its significance, policy-makers, political leaders and opinion-makers to become more active to warn about its consequences. The significant asymmetric impact favouring easier as opposed to tighter monetary policy suggests that the aggressive unconventional monetary policy pursued by the FED from 2008 worked in the desired direction.\\

%\textcolor[rgb]{1.00,0.00,0.00}{From 2008, expected inflation became persistently higher than realized inflation. News of unconventional monetary policy acts as an imperfect signalling device for household inflation expectations. Households lost faith to some extent about the effectiveness of policy in this period, which raised inflation expectations well above the eventually realized inflation, which fortuitously meant lower real interest rates.  Thus we find irrelevance for expected inflation of conventional monetary policy at the zero lower bound of the federal funds rate. Forward guidance was one aspect of the unconventional monetary policy employed, and we find evidence that it acted as a signalling device for households about future inflation. Importantly, with the disconnect of conventional monetary policy news, weak consumer sentiment through perceived credit market conditions played an important role in understanding the relatively high inflation expectations in this period.}\\

From 2008, expected inflation became persistently higher than realized inflation.  We find news of unconventional monetary policy acts as an imperfect signalling device for household inflation expectations, which may be seen as a proxy for consumer sentiment in relation to inflation expectation formation. Weak consumer sentiment through perceived credit market conditions may have played an important role in understanding the relatively high inflation expectations in this period.\\

%%%%%%%%%%%%%%%%%%%%%%%%%%%%%%%%%%%%%%%%%%%%%%%%%%%%%%%%%%%%%%%%%%%%%%%%%%%%%%%%%%%%%%%%%%%%%%%%%%%%%%%%%%%%%%%%%%%%
%\newpage
%\newgeometry{top = 2.0cm, left = 1.5cm, right =1.5cm}
%\section{The role of the size of the estimated measurement error variances}  \label{sec:BCI_Appendix_merr}
%\restoregeometry

%%%%%%%%%%%%%%%%%%%%%%%%%%%%%%%%%%%%%%%%%%%%%%%%%%%%%%%%%%%%%%%%%%%%%%%%%%%%%%%%%%%%%%%%%%%%%%%%%%%%%%%%%%%%%%%%%%%%%%%%%%%%%%%%%%%%%%%%%%%%%%%%%%%%%%%%%%%%%%%%%%%%%%%%%%%%%%%%%%%%%%%%%%%%%%%%%%%%%%%%%%%%%%%%%%%%%%%%%%%%%%%%%%%%%%%%
%%%%%%%%%%%%%%%%%%%%%%%%%%%%%%%%%%%%%%%%%%%%%%%%%%%%%%%%%%%%%%%%%%%%%%%%%%%%%%%%%%%%%%%%%%%%%%%%%%%%%%%%%%%%%%%%%%%%
%\newpage
%\section{Autocorrelations}

%%%%%%%%%%%%%%%%%%%%%%%%%%%%%%%%%%%%%%%%%%%%%%%%%%%%%%%%%%%%%%%%%%%%%%%%%%%%%%%%%%%%%%%%%%%%%%%%%%%%%%%%%%%%%%%%%%%%%%%%%%%%%%%%%%%%%%%%%%%%%%%%%%%%%%%%%%%%%%%%%%%%%%%%%%%%%%%%%%%%%%%%%%%%%%%%%%%%%%%%%%%%%%%%%%%%%%%%%%%%%%%%%%%%%%%%
%%%%%%%%%%%%%%%%%%%%%%%%%%%%%%%%%%%%%%%%%%%%%%%%%%%%%%%%%%%%%%%%%%%%%%%%%%%%%%%%%%%%%%%%%%%%%%%%%%%%%%%%%%%%%%%%%%%%

%\clearpage
%\setlength{\textheight}{592pt}        %%default 592pt
%\setlength{\textwidth}{390pt}         %%default 390pt
%\setlength{\marginparwidth}{35pt}      %%default 35pt
%\setlength{\hoffset}{0pt}
%\setlength{\voffset}{0pt}     %%default 0
%\setlength{\footskip}{0cm}       %%default 0
%\setlength{\parskip}{0cm}	
%\setlength\arraycolsep{9pt}

\newpage

\section{Reference}
%\bibliography{../Ref}
%\bibliographystyle{elsart-harv}

\newpage
\appendix
\section{Pairwise correlation}   \label{App:Corr}
    \begin{figure}[h]
      \begin{center}
        \includegraphics[scale=0.95,angle=0]{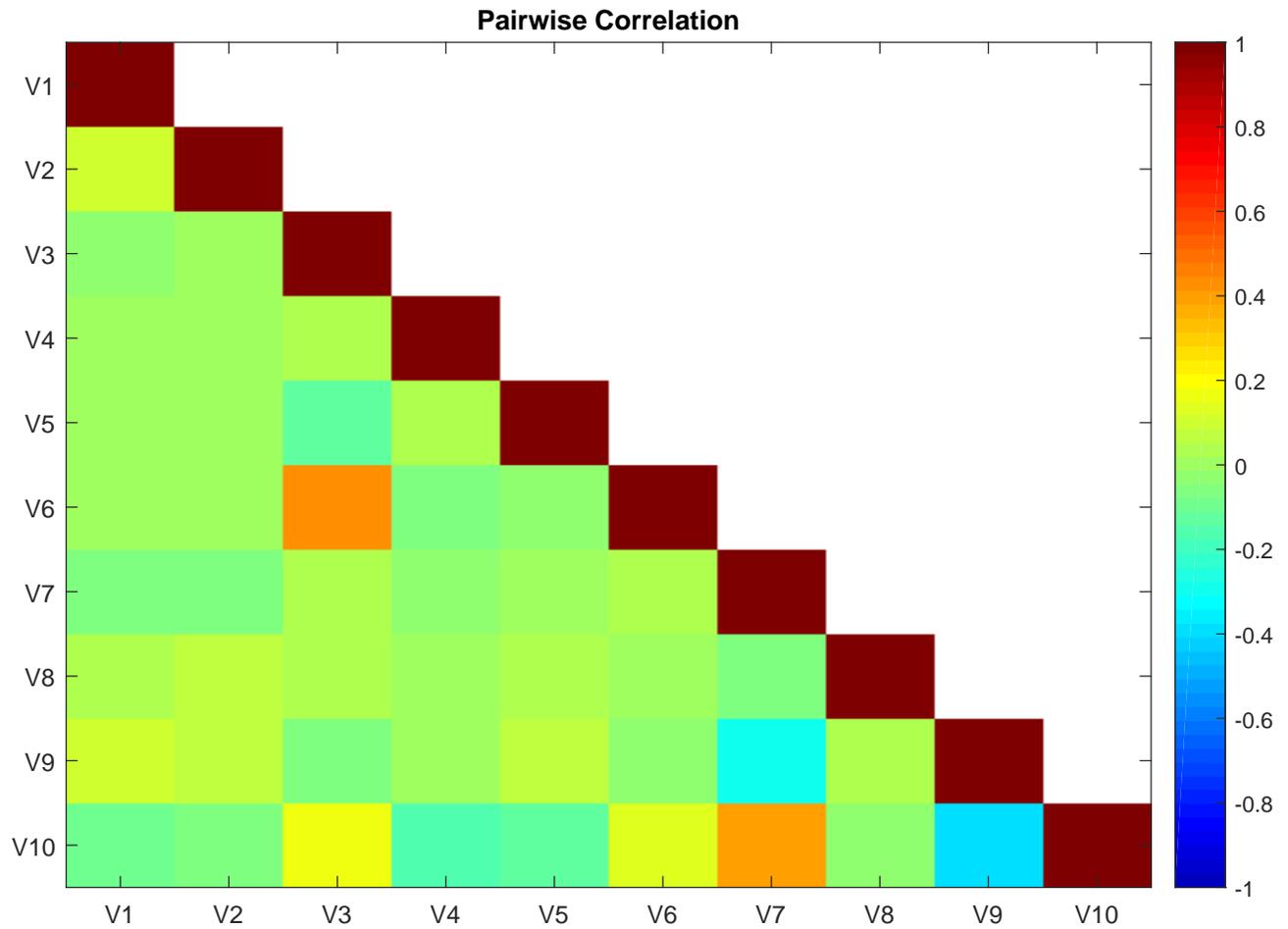}
        \caption{V1: News: inflation; V2: News: monetary policy; V3: Log income; V4: Age; V5: Female; V6: Education; V7: Perception: government policy; V8: Expectation: interest rate; V9: Expectation: unemployment rate; V10: Consumer sentiment;}
        \label{fig:Xcorr}
     \end{center}
    \end{figure}
\end{document}